# Intermediate diffusive-ballistic electron conduction around mesoscopic defects in graphene

*Toni Markovic,[†,‡] Wei Huang,[†,‡] William S. Huxter,[§] Pietro Gambardella,[†] Sebastian Stepanow[†],\**

[†]Department of Materials, ETH Zurich, Honggerbergring 64, 8093 Zurich, Switzerland

[§] Department of Physics, ETH Zurich, Otto-Stern-Weg 1, 8093 Zurich, Switzerland

*Corresponding author e-mail address: sebastian.stepanow@mat.ethz.ch

**Abstract**

Non-diffusive effects in charge transport become relevant as device sizes and features become comparable to the electronic mean free path. As a model system, we investigate the electric transport around mesoscopic defects in graphene with scanning tunneling potentiometry. Diffusive and ballistic contributions to the scattering dipole are probed by simultaneously resolving the nanoscale topography of pits in the graphene layer and measuring the local electrochemical potential in the surrounding area. We find evidence of transport in the intermediate regime between the diffusive and ballistic limits, such that the magnitude of the electrochemical potential around the defects is substantially underestimated by diffusive models. Our experiments and modeling are supported by lattice Boltzmann simulations, which highlight the importance of the ratio between defect size and mean free path in the intermediate transport regime. The magnitude of the scattering dipole depends on the shape of the pits in both the ballistic and diffusive transport modes. Remarkably, ballistic contributions to the electron transport are found at feature sizes larger than the mean free path and rapidly increase at lower sizes, having a noticeable impact already at mesoscopic length scales.

The crossover from diffusive to ballistic charge transport is a regime whose importance increases as device sizes and dimensionality continue to decrease.[1–7] The characteristics of the intermediate transport regime, where device dimensions are comparable to the electron mean free path, have been addressed theoretically using semi-classical[7–11] or quantum transport approaches.[12–14] On the experimental side, the typical top-down approaches that aim to elucidate emergent ballistic effects are hindered by the lack of local and spatial information and electrical measurement signals are obscured by scattering at device contacts and edges.[5,15–17] Thus, experimental investigations in this transport regime would strongly benefit from local measurements at nanometer length scales, ideally with non-invasive probes that facilitate simple interpretations and eliminate potential artifacts. Moreover, there is no systematic local study available addressing the diffusive-ballistic intermediate transport regime that allows for a direct comparison with predictions from theory. Several scanning probe techniques such as nitrogen-vacancy microscopy, scanning superconducting quantum interference devices, and single-electron transistors offer spatial resolutions with 10's of nanometers[18–22] and have had some success in imaging ballistic transport features.[23] One technique with sub-nanometer resolution is scanning tunneling potentiometry (STP), shown in Fig. 1a, which simultaneously measures topography and current flow through the local electrochemical potential (ECP).[24–26]

Recent studies with STP have led to exciting insights concerning transport in and around graphene nanostructures,[27–35] including ballistic transport in nanoribbons[36] and viscous effects through potential constrictions.[37] Graphene is an ideal material platform for such studies as it has a relatively long mean free path, confines current flow into a two-dimensional (2D) plane, and provides numerous geometries to explore. Transport across the monolayer-bilayer boundary (Fig. 1b,c), for example, features an abrupt resistance change accompanied by a narrow transition region.[27,30–32] The spatial extent of this transition region has been tied to the Landauer residual resistivity dipole (RRD), which characterizes the ECP profile due to ballistic scattering at a point-like defect.[38,39] More generally, the RRD is appropriate for any scatterer that is small compared to the electronic mean free path (Fig. 1d).[40] Its observation is usually considered a clear indication of ballistic contributions to transport, however, similar ECP profiles can be reproduced using diffusive transport models.[41] Importantly, diffusive and ballistic ECP profiles distinguish themselves in the magnitude of the resistivity dipole. In addition, it is expected that the two transport modes exhibit a different dimensional dependence of the scatterer, i.e., diffusive ECP profiles scale with the scatterer area ($\sim L^2$) whereas ballistic ECP profiles scale with the scattering cross section ($\sim L$).[42] Little is known regarding ECP profiles in the intermediate regime, where the mean free path approaches the size of the scatterer.

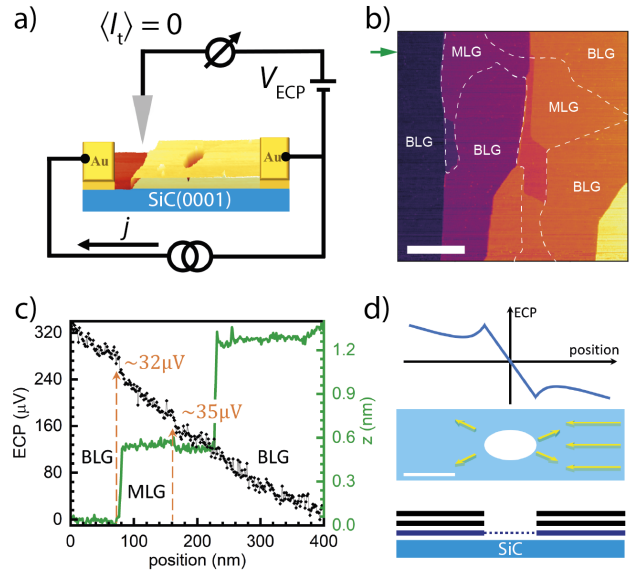

**Figure 1**: (a) Schematic of scanning tunneling potentiometry. An external current source supplies current through metal contacts. A scanning tunneling microscopy tip is brought a few Angstroms away from the surface and measures the tunneling current. The local ECP is measured with µV resolution (simultaneously with the topography) by applying an additional voltage to the graphene. The voltage is equal to the electrochemical potential when the tunneling current is nulled. (b) Topography of graphene/SiC with a mixture of monolayer and bilayer graphene terraces on SiC triple steps. The separation between the monolayer and bilayer graphene is highlighted with dashed white lines. Scale bar is 100 nm. (c) The 5-line averaged topography (green) and ECP profile (black) acquired across the line indicated by the green arrow in (b). Two monolayer-bilayer interfaces are indicated by voltage drops (orange arrows). (d) Upper part: Sketch illustrating the response of the local ECP in the presence of a localized scatterer. The yellow arrows represent the direction of the carriers, and the white bar represents the electron mean free path. Lower part: Sketch of the cross section of the sample. The SiC substrate and the graphene sheets (black) are separated by a carbon buffer layer (dark blue). Within the pit, there is amorphous carbon instead of the buffer layer.

Here, we use STP to investigate charge transport in graphene around naturally occurring voids, or pits, that have atomically sharp boundaries and act as scatterers with sizes between 10 and 100 nm, a similar order to the electron mean free path.[43] Our STP data reveal ECP dipole profiles that are substantially stronger than expected for diffusive transport. Combining diffusive and ballistic models with lattice Boltzmann simulations, we reveal a clear non-diffusive character which we attribute to ballistic effects in the intermediate regime. Collectively, our data and simulations demonstrate that the onset of ballistics effects become important even at scatterer sizes significantly larger than the electron mean free path. Moreover, geometric factors (for scatterers deviating from an ideal circular shape) also need to be considered to properly quantify ballistic and diffusive influences on the transport.

**Results**

**Imaging charge transport in the intermediate regime**

Figure 1b provides an overview of the sample topography. The larger height variations ($\sim 0.5 - 0.8$ nm) are caused by multiple SiC substrate steps. On the surface, we observe regions of monolayer graphene (MLG) and bilayer graphene (BLG), which can be distinguished by their topographic features, as well as by signatures in d$I$/d$V$ spectroscopy and thermal voltage from STP measurements. The slight height difference between MLG and BLG on the same terrace arises from a lower SiC substrate step beneath the BLG.[25] The pits are formed[44–46] during the epitaxial growth of monolayer graphene (MLG) and bilayer graphene (BLG) on SiC via thermal decomposition.[47–49] Most pits are embedded in BLG and some have lateral interfaces to both MLG and BLG. We measure at 295 and 90 K to study the influence of temperature. STP data for two typical pits (at low and high temperatures) are summarized in Fig. 2. The topographs in Fig. 2a,f display the easily identifiable pits, with a typical 600 pm depth and corrugated interior. Their depth corresponds to the BLG height (~ 660 pm) above the so-called buffer layer – a graphene-like arrangement of covalently bonded carbon on the SiC substrate. Since we do not observe the characteristic $6\sqrt{3} \times 6\sqrt{3}$ reconstruction of the buffer layer inside the pit (see Supplementary Note 1), we conclude that the pit consists of a non-reconstructed amorphous carbon layer interrupting the graphene sheet.[50,51]

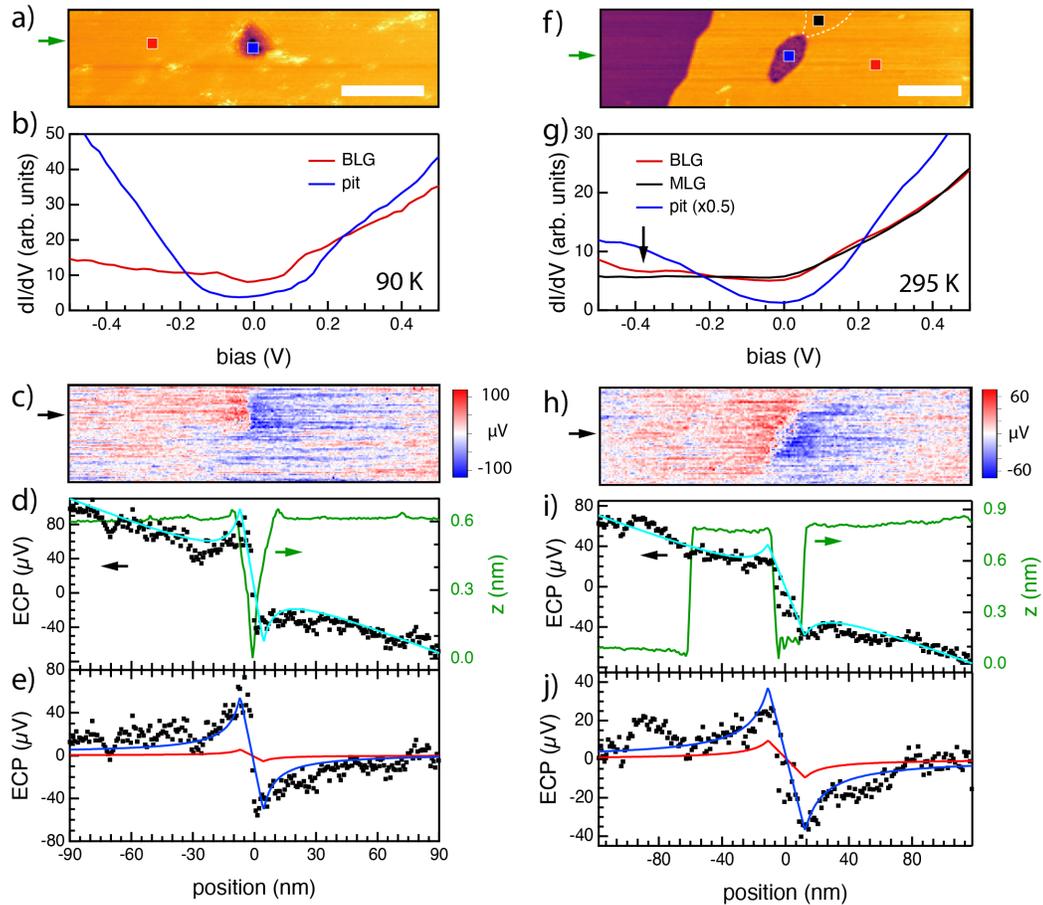

**Figure 2:** STP data for selected pits obtained at (a-e) 90 K and (f-j) 295 K. (a,f) Topographs show pits in the graphene film. The pits (marked with blue squares) are surrounded by BLG (red squares)

and a small region of MLG (black square, within the dashed lines) in panel (f). (b,g) d$I$/d$V$ spectra obtained at the positions indicated with correspondingly colored symbols in (a,f) showing characteristic features used for surface classification. Vertical arrow in (g) marks the Dirac point for BLG. The dI/dV data were obtained by averaging 20 curves with an acquisition time of 10 ms per point. (c,h) Plane subtracted ECP maps of scan areas (a,f), respectively. (d,i) Cross sections of the topography and ECP map at the arrows indicated in panels (a,c,f,h). Five adjacent scan lines taken across the center of the pit were averaged (see Supplementary Note 1 for more details on the fitting procedure and residual noise of the data as represented by the standard deviation of the five lines averaged). The solid cyan line represents a fit (see text). The black and green arrows indicate the corresponding y-axis for the plotted data. (e,j) ECP data after a linear background subtraction. Solid lines represent potential profiles using diffusive (red) and ballistic (blue) models (see text) and are not fits to the data. STM setpoint: (a) 100 mV bias voltage and 600 pA tunneling current, (f) 50 mV bias voltage and 500 pA tunneling current. Applied bias current density across the sample is 4 A/m. Scale bars in (a,f), 40 nm.

To characterize the electronic properties of the pits, we performed scanning tunneling spectroscopy (STS), as shown in Fig. 2b,g. The data show that the BLG has a characteristic dip at −380 mV in the d$I$/d$V$ spectrum, indicating the Dirac point, whereas the MLG spectrum is flat around the same value, as expected from previous reports.[47–49] The d$I$/d$V$ spectrum acquired within the pit resembles that of the buffer layer and indicates substantially less conductive behavior at the Fermi level compared to graphene.[47,48,52] STS measurements also show a sharp change across the pit to MLG or BLG (see Supplementary Note 1 for further analysis), demonstrating that the pit is a sharp discontinuity both topographically and electronically. Furthermore, the electronic properties of the graphene sheet are not significantly perturbed near the boundary of the pit.

To investigate the pit's influence on transport, we apply a bias current through the graphene (along the horizontal $x$-direction in Fig. 2a,f) and measure the ECP. The ECP is shown in Fig. 2c,h after subtraction of a plane. This procedure removes the slope stemming from the temperature-dependent resistivity from the ECP data and allows for a quantitative analysis of the scattering profile (Supplementary Note 1). The dipolar structure – an increased ECP on one side and a decreased ECP on the other side – is clearly visible in the data. Traces of the original ECP profiles (marked by an arrow in Fig. 2c,h) are shown in Fig. 2d,i, where five adjacent scan lines have been averaged to reduce noise. Comparing the cross-sections of the topography and ECP (cf. Fig. 2d,i), we see that the positions of the pit edges and the peaks in the ECP match exactly.

**Modeling the transport dipole**

To analyze the ECP profiles, we derive diffusive and ballistic models for circular pit shapes and compare their predictions to our measurements. The circular pit serves as a general reference for analyzing the different transport regimes. The influence of the pit shape on the dipole magnitude is discussed in the following sections. For the diffusive model, we use Ohm's law and calculate the potential around a circular hole in a 2D sheet of homogeneous resistivity in a constant electric field. The solution of the Laplace equation satisfying the boundary conditions is the dipolar potential[53]

$$V_{dip}^d = -\rho j \frac{a^2}{r} \cos \phi, \qquad (1)$$

where $\rho$ is the sheet resistance, $j$ is the 2D current density far away from the hole, $a$ is the hole radius, $\phi$ is the azimuthal angle taken from the hole center and $r = \sqrt{x^2 + y^2}$. For the ballistic model, we use the Landauer formula[3,38] for a scatterer in 2D and calculate the transport cross-section for a circular scatterer of radius $a$,

$$V_{dip}^b = -\frac{16\hbar}{3k_F e^2} j \frac{a}{r} \cos\phi, \tag{2}$$

where $k_F = 0.72$ nm$^{-1}$ is the Fermi wavevector for the BLG/SiC system[49,54] and $e$ is the elementary charge. With $\frac{1}{\rho} = \frac{2e^2}{h} k_F l$, where $l$ is the electron mean free path, the formula can be rewritten as

$$V_{dip}^b = -\rho j \frac{16}{3\pi} l \frac{a}{r} \cos\phi. \tag{3}$$

Both models are nearly identical in functional form, however, the ballistic model scales linearly with the pit radius $a$ whereas the diffusive model scales quadratically with $a$.

To combine both models in our analysis, we define a dipole magnitude $M$ (with units of nanometers) using a general dipole potential along the current direction $x$ at the center of the pit,

$$V_{dip}(x) = \text{sgn}(x) \rho j \frac{a}{x} M, \quad |x| \geq a. \tag{4}$$

The dipole magnitude $M$ has the advantage that it does not depend on the local sheet resistance or current density. To extract $M$, we normalize the data by the driving field $\mathcal{E} = \Delta V / \Delta x = \rho j$, which is directly obtained from the slope of the ECP plane subtraction. The dipole magnitude $M$ is essentially the maximum value of $V_{dip}(x = a)/\mathcal{E}$ at the edge of the pit. Importantly, $M$ scales linearly with the pit radius in the diffusive case ($M_d = a$) but is independent on the pit size in the ballistic case ($M_b = \frac{16}{3\pi} l$), depending only on the mean free path. Evaluating the mean free path to $l = 33 \pm 5$ nm at 90 K (Supplementary Note 1), we obtain $M_b = 54$ nm, which yields a rough estimate for the ballistic dipole magnitude. Thus, measurements of $M$ serve as a direct signature of diffusive and ballistic effects on the transport.[50,51] At 295 K, the mean free path cannot be directly determined from the sheet resistance measurements because part of the current flows through the SiC substrate. Instead, we use the mean free path obtained at 90 K as an upper limit for the 295 K data. This approach aligns with measurements showing that the electron mobility in BLG is only weakly temperature-dependent in this range, particularly at the high doping levels observed for SiC.[55,56] For comparison, data for MLG indicates that the mean free path decreases by roughly a factor of 1.5 between 80 K to 300 K.[29,43] Additionally, the normalization procedure of the dipole potential remains unaffected, as $\Delta V/\Delta x = \rho_{total} j_{total} = \rho_{graphene} j_{graphene}$ due to the parallel resistances of graphene and SiC.

We fit the data in Fig. 2d,i using Eq. (4), adding a linear slope for the sheet resistance, and find good agreement. Figure 2e,j shows the plane subtracted ECP profiles. The dipole magnitude of the low temperature data ($M = 87 \pm 4$ nm) is larger than the high temperature data ($54 \pm 3$ nm), however, the dimensions of the pits in the two cases are also different, namely 12 nm along $x$ for the pit shown in Fig. 2a and 30 nm for the pit measured in Fig. 2f. Both dipole magnitudes lie closer to the estimated ballistic value than the expected diffusive values, which we estimate as 6 and 15 nm, respectively, from half of the pit length in the $x$ direction ($M_d = a$). Note that these estimates for the diffusive dipole magnitudes carry uncertainty, which arises not only from the ambiguity in the pit boundary due to topography and dipole fitting but also from deviations in pit shape from a perfect circle. For comparison, we also plot the profiles for the diffusive and ballistic model, which supports our finding of a substantial ballistic contribution. However, the transport is not purely ballistic, as evidenced by the finite slope of the ECP. Moreover, the relatively large dipole magnitudes observed at 295 K are notable for several reasons. First, the mean free path is expected to decrease compared to 90 K, suggesting that the ballistic limit assumed for the circular shape in Fig. 3b is an overestimate. Second, the current density in graphene is reduced due to parallel transport through the SiC substrate (as discussed in Supplementary Note 1). While this is accounted for by the normalization procedure, the behavior of the scattered electrons at the pit boundaries is more complex. In principle, electrons could scatter into the SiC substrate, which would reduce the dipole magnitude. However, we anticipate minimal coupling between the graphene sheets and the SiC substrate due to passivation of the SiC by the buffer layer.

Some deviation from the $1/r$ dipolar extension of the profiles is noticeable, which we partially ascribe to the surrounding topographic inhomogeneity as well as the shape of the pit since both models assume a perfect and impenetrable circle in an otherwise homogeneous conducting sheet. Furthermore, the potential profile in the ballistic case can deviate from the dipolar potential close to the pit.[3,38] In addition, although the pit is a discontinuity in the graphene, a finite conductivity persists inside the pit (as shown by STM and STS), i.e., not all the incoming electrons are backscattered. Therefore, our modeling is only approximate, but nonetheless provides useful boundaries to define the intermediate regime between diffusive and ballistic transport.

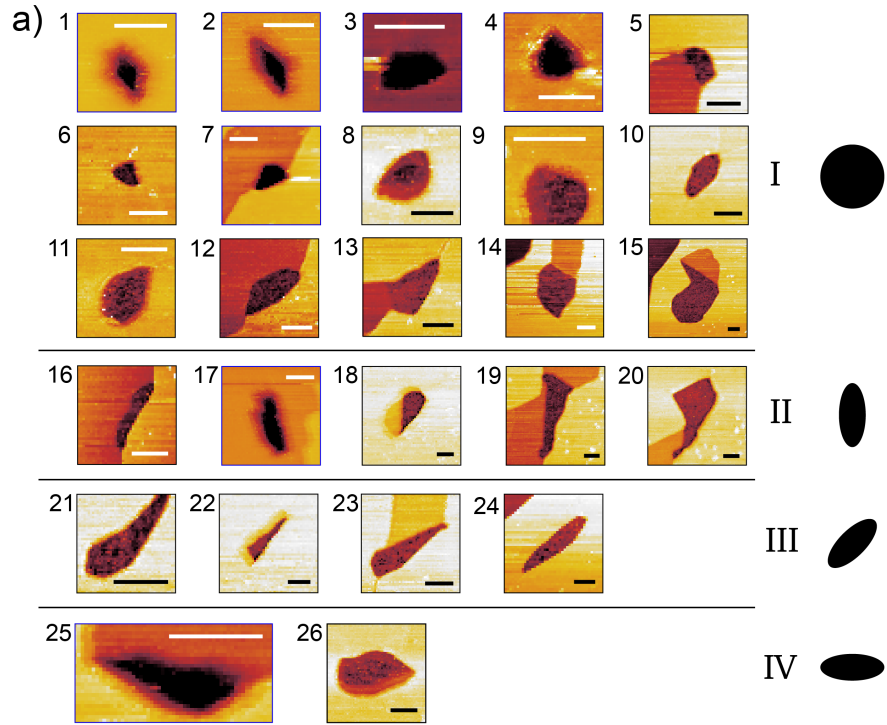

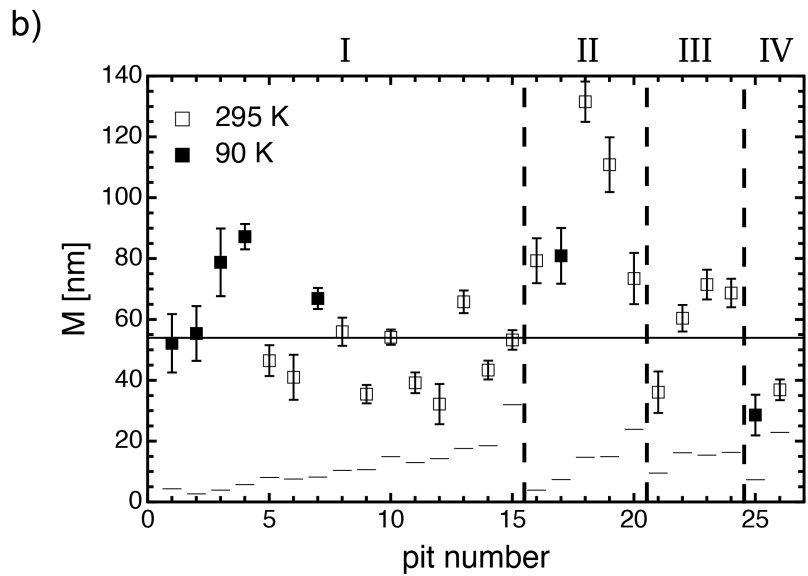

**Figure 3:** (a) STM topographies of the investigated pits at 90 and 295 K with the current applied in the horizontal direction. The pits have been grouped by visual inspection of their shape into four categories: (I) round, sometimes with a tail, (II) needle-like, with the long extension perpendicular to the current, (III) needle-like, oriented at ~45° with respect to the current, (IV) needle-like, with the long extension parallel to the current. Scale bar is 20 nm in all images. (b) Fitted and normalized experimental dipole magnitudes at 295 K (open squares) and 90 K (filled squares) for the pits numbered in (a). The continuous horizontal line and the horizontal line markers represent the values expected for a circular pit of the same length along $x$ from the ballistic and diffusive model, respectively. Within each category, the pits are ordered by their length along $x$.

**Pit-shape anisotropy**

To account for pit size and shape, we acquired ECP maps across 24 additional pits (Fig. 3a), at 295 K and 90 K, with a broad range of dimensions. The corresponding dipole magnitudes are presented in Fig. 3b, along with the prediction from the diffusive and the ballistic models. For the diffusive model, we again estimate $M_d = L_x/2$, where $L_x$ is the horizontal width and taken roughly in the vertical middle of the pit. As noted above, this represents an approximate estimate for the expected dipole magnitude from the diffusive model. We additionally divide the pits into four categories (labelled I-IV in Fig. 3b), based on their overall shape, and sort them by $L_x$ within each category.

The experimental values of $M$ lie substantially above the diffusive model and their spread is relatively large, from the diffusive boundary to well beyond the ballistic prediction. Comparing the data for 90 K and 295 K, we do not observe a significant difference, which can be attributed to the similar mean free paths in this temperature range (see supplementary information for a more analysis of the mean free path). For round pits (category I, $L_x \approx L_y$, where $L_y$ is the cross section perpendicular to the current) we find values in the whole range between diffusive and ballistic models. The data does not reveal any significant size dependence of $M$. Additionally, while we would expect that the dipole magnitude for larger pits approaches the diffusive limit (particularly for rounds pits), the largest pit (no. 15, $L_x/2 \approx 32$ nm) has a measured dipole magnitude ($M = 53$ nm) well above the diffusive prediction. This strongly suggests that we are positioned in the intermediate transport regime.

The observed spread of dipole magnitudes also appears to be strongly influenced by pit shape, beyond the amount of variation observed for circularly-shaped pits. Supported by mesoscopic simulations detailed below, we interpret this increased spread as a combination of ballistic effects, as the cross section of the pits across categories II-IV vary, and a geometric factor. For needle-like pits perpendicular to the current direction (category II, $L_y > L_x$), we find dipole magnitudes close to and beyond the ballistic limit. If the needle-like pits have an angle at about 45° with respect to the current (category III), their dipole magnitudes drop slightly. If the elongated pits are parallel to the current direction (category IV, $L_y < L_x$), the dipole magnitudes fall closer to the diffusive regime since their perpendicular cross section is small compared to their overall size.

**Mesoscopic transport simulations**

To explore the intermediate transport regime and characterize the shape dependence, we developed a mesoscopic transport model using the Lattice Boltzmann Method (LBM), tailored for a classical electron fluid[57] (see Supplementary Note 2). The LBM is a simulation framework based on the Boltzmann transport equation that iteratively solves for a set of steady-state quasi-particle density functions on discrete lattice points that define a simulation geometry. At each lattice point the quasi-particle density function is broken into discrete energy- and velocity-dependent density components which redistribute across neighboring lattice points in alternating streaming and collision processes. The LBM, traditionally used for hydrodynamics,[58] offers great flexibility for simulating arbitrary geometries. We incorporated diffusive scattering by adding momentum relaxation in the collision process and simulated charge transport around specularly reflecting pits. By appropriately setting the momentum-relaxing and momentum-conserving lengths ($l_{mr}$ and $l_{ee}$, respectively), we can tune transport from the diffusive limit towards the ballistic regime. In these simulations $l_{mr}$ is equivalent to the electronic mean free path (Supplementary Fig. S8) and $l_{ee}$ captures hydrodynamic interactions, which do not play a significant role in this analysis (Supplementary Fig. S9).

Across the simulation geometry, the steady-state quasi-particle densities can be used to compute spatial quantities, such as the current density and the local ECP. An example of a plane subtracted ECP with a clear dipole shape, is shown in Fig. 4a. The pit radius $a = 40$ nm is substantially larger than $l_{mr} = 5$ nm and diffusive interactions dominate the transport. When $l_{mr}$ is increased to 40 nm (Fig. 4b, better reflecting experimental values), ballistic contributions are visible through an increased dipole magnitude and appearance of ray like structures. It is important to note that the discrete velocity scheme used in our model produce small scattering artifacts (ray effects) when ballistic effects dominate the transport; in reality, reflections at pit edge should be radially symmetric. To account for this well-established problem,[59] we limit ourselves to discussions of weak ballistic effects, i.e., to a regime where $l_{mr} \leq 2L_x, 2L_y$.

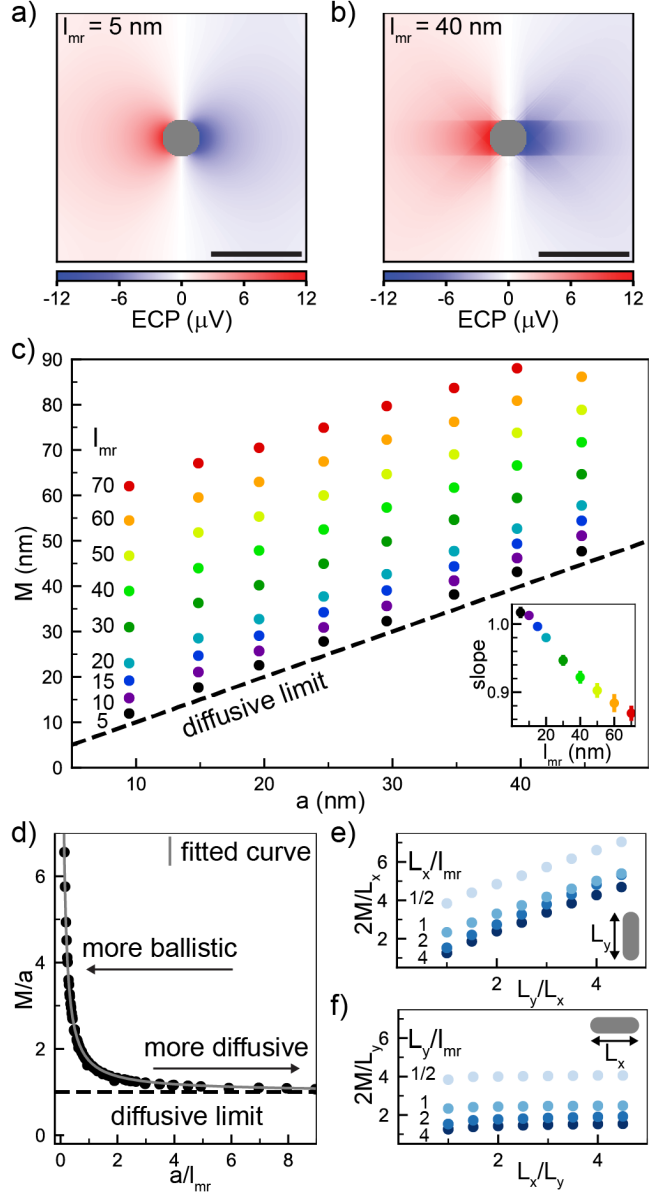

**Figure 4:** (a) Plane subtracted ECP map computed from the lattice Boltzmann method. A circular pit of radius $a = 40$ nm is defined inside the BLG sheet and $l_{mr} = 5$ nm. (b) Same as (a) with $l_{mr} = 40$ nm, representing a realistic scenario from experiments. (c) Dipole magnitude $M$ for different pit radii $a$ and $l_{mr}$. The diffusive model value is also shown with a dashed black line. The inset plots the slope of $M$ as a function $a$ for different $l_{mr}$. A slope of one (zero) corresponds to the diffusive (ballistic) transport limit. The slopes slightly above one are a result of discrete

circles constructed on a square grid. (d) All values of $M/a$ align to a single curve when plotted as a function of $a/l_{mr}$. The gray curve is a fit of $M/a = 1 + c/(a/l_{mr})$, where $c$ is a unitless constant. (e) Effect of extending $L_y$ for different $l_{mr}$ and aspect ratios. Normalization of $M$ by $L_x$ allows for a size-independent interpretation of the results as is done in (d). (f) Same as in (e) except $L_x$ is varied and the values are normalized by $L_y$. In all simulations, $l_{ee} = 500$ nm. Scale bar in (a) and (b), 200 nm.

By systematically varying $a$ and $l_{mr}$ we can compare the simulated dipole magnitudes, shown in Fig. 4c, to experiments. $M$ is extracted along the symmetry axis of the pit along the $x$-direction (after plane subtraction). For very small $l_{mr}$, we recover the diffusive limit and $M$ is approximately equal to the pit radius. However, as $l_{mr}$ increases, $M$ increases beyond the diffusive limit nearly proportional to $l_{mr}$, similar to our experimental observations and theoretical modeling. The simulations also show a pronounced dependence of $M$ on the pit radius, even for $l_{mr} > a$. By fitting the slope of $M$ as a function of $a$ (inset in Fig. 4c) we quantify ballistic contributions on the transport dipole and observe that the slope decreases below a value of one, where a slope of one corresponds to pure diffusive transport. To characterize the increase in $M$ from ballistic interactions, we plot the dipole magnitude normalized by the pit radius as a function of the ratio between the pit radius and momentum-relaxing length. When $a/l_{mr}$ is large we expect $M/a$ to be equal to one, corresponding to the diffusive limit. As $a/l_{mr}$ decreases, ballistic contributions will increase $M$ beyond $a$. We find that all simulated data follow a single curve (Fig. 4d) that is inversely proportional to $a/l_{mr}$, capturing the smooth transition from diffusive transport into the intermediate regime. Note that such a scaling behavior is analogous to the scaling of the conductance across the intermediate regime, $G \sim \left(1 + c'\frac{L}{l}\right)^{-1}$,[7–9,36] where the resistance increases when the transport is more diffusive. Contrastingly for our pit geometry, the dipole magnitude increases with ballistic contributions. Thus, the quantity $M/a$ can be seen as a direct measure of the ballisticity of the transport.

Motivated by the large variability in $M$ for pits in categories II-IV, we also simulate pits with different aspect ratios to investigate any geometrical effects. Extending the pit shape perpendicular to the current flow (Fig. 4e) results in an increased dipole magnitude. This contribution is different than those from ballistic effects and is present across all simulations from the diffusive to the ballistic regime. Extending the pit parallel to the current flow (Fig. 4f), in comparison, affects the overall dipole magnitude very little. With this data, we can partially explain the shape dependence observed in Fig. 3b. These simulations suggest that pits belonging to category II are subject to an additional geometric effect that increases the observed dipole magnitude beyond the ballistic value. This geometric effect is reduced as the needle-like pits rotate parallel to the current flow (categories III and IV).

**Discussion**

Our experimental measurements, theoretical modeling, and transport simulations clearly show signatures of ballistic effects in the transport of MLG and BLG on SiC. Such an intermediate transport regime is consistent with the mean free path (which is about 30 nm) being comparable to the dimensions of the pits. Other ECP dipole profiles have also been observed around pits in thin metallic films[60,61] and in topological insulators,[62] where the electron mean free path amounts to only a few nanometers and pit diameters are around 10−20 nm. In the topological insulator case, a good agreement with a resistor network model was found. For the Bi films, the analytical diffusive dipole model accounted for most of the measured dipole magnitude and a small ballistic contribution was suggested by the authors. Our data and simulations suggest that ballistic effects add substantially to the dipole magnitude already for pit dimensions significantly larger than the electron mean free path, and that geometric effects should also be considered when the pit is elongated perpendicular to the current flow equally for both transport regimes.

Apart from an increased dipole magnitude, we did not observe additional evidence for ballistic wave-like transport, which should give rise to Friedel-like oscillations in the dipolar potential.[3,4,39] These oscillations are very difficult to observe as they would add a small correction on top of the dipolar ECP profile. Given the small electronic wavelength at our relatively high Fermi energy (high intrinsic doping) and the irregular shapes of the pits, those effects are most likely also smeared out in the long-range ECP profiles. We anticipate that studies with Fermi level manipulation, either through intercalation[63] or with gated substrates,[37] will enable direct access to measure the influence of the electron wavelength and mean free path in the intermediate transport regime.

Thus far, we have disregarded the electron-electron interaction length $l_{ee}$, which can play a significant role in graphene transport. Since hydrodynamic effects on transport in graphene have been demonstrated exclusively in extremely pure samples with special geometries,[64] we do not expect it to play a role for our pit measurements.[65,66] We also show in the SI that the trends observed in our LBM simulations are independent of $l_{ee}$ across two orders of magnitude.

**Conclusions**

In summary, we demonstrate that naturally occurring pits in graphene present a strong scattering center in the intermediate regime between diffusive and ballistic transport. The experimentally observed magnitudes of the resistivity dipoles substantially exceed those predicted by diffusive transport already at pit dimensions beyond the electron mean free path. We also show that shape effects must also be considered since they appear in both transport regimes equally and could potentially only be attributed to ballistic effects. The onset of ballistic effects at feature sizes larger than the mean free path and their rapidly increasing predominance at lower sizes is of great relevance for nanoscale device applications.

**Methods**

**Experimental methods**

The graphene was epitaxially grown on a commercial SiC-6H semi-insulating substrate (MSE supplies, thickness 330 μm, resistivity > $1 \times 10^5$ Ωcm) by thermal decomposition in ultra-high vacuum ($1 \times 10^{-10}$ mbar). The SiC-6H wafer of dimension $8 \times 0.5$ mm$^2$ was directly heated by passing an electrical current through the sample, while its temperature was monitored with a pyrometer (emissivity set to 0.75 at wavelength 1.2 μm). The SiC was heated this way for 10 cycles going from 600 °C to 1400 °C at a rate of 10 °C/s, where it is kept for 30 s before cooling it back down with the same rate. We used slow heating and cooling rates in order to promote step bunching to create wider terraces and promote pit formation on the sample surface.[44–46] We kept the number of cycles low in order to keep the graphene layer count below two. The sample was predominantly covered in bilayer graphene with smaller areas covered in monolayer graphene. Sample contacts were made ex-situ by pressing Indium foil with a thickness of 127 μm directly onto the graphene by using Ta clamps (295 K data) or by depositing Au contacts (thickness 50-100 nm) using a shadow mask (90 K data). These methods created a good contact resistance (∼100 Ω). After reintroducing the sample into ultra-high vacuum, it was degassed at about 100 °C for 30 min to remove adsorbates.

We have measured ECP by using our dc mode STP implementation based on a point-by-point current direction switching method, as explained in Ref. 25, to remove artifacts from the thermal voltage in the ECP measurements. Our STP implementation is based on the sample-and-hold approach, i.e., the *z*-topography feedback is frozen to keep the tip-sample separation constant while another feedback adjusts the compensation voltage such that the tunneling current is nulled. At that point the tip and sample at the position of the tip are equipotential, making the compensation voltage equal to the ECP of the surface. The measurements were performed at room temperature with an applied 4 A/m current density. Typically, we have used 100 mV and 500 pA as the scanning parameters and a 200 ms integration time per point for the ECP acquisition.

**Supporting Information Available**.

STM and STS characterization of the graphene pit, Determination of the dipole magnitude, Sheet resistance measurements of BLG/SiC, Temperature-dependent resistivity and estimation of the mean free path in BLG; Lattice Boltzmann method: Description and model validation, Effect of boundary scattering on pit geometry, Resistivity and momentum relaxation, Investigation of hydrodynamic effects around pits; Topography, ECP maps and traces for all pits

**Data availability**. The data that support the findings of this study are available from the ETH Research Collection under doi:10.3929/ethz-b-000724858

**Author Contributions**

‡These authors contributed equally. T.M. and W.H. performed the measurements. T.M., W.H. and S.S. designed the experiments and analyzed the data. T.M. developed the models for diffusive and ballistic dipole magnitudes. W.S.H. wrote the code for the LBM, carried out the numerical simulations and the subsequent analysis. All authors discussed the results and prepared the manuscript.

**Funding Sources**

Swiss National Science Foundation, Project No. 200021_163225 and 200020_207478.

China Scholarship Council (CSC) under Grant No. 201806340092.

European Research Council through ERC CoG 817720 (IMAGINE).

**Acknowledgement**

The authors thank M.L. Palm and the Euler computer cluster at ETH Zurich. This work was supported by the Swiss National Science Foundation, Project No. 200021_163225 and 200020_207478. W.H. acknowledges supported by the China Scholarship Council (CSC) under Grant No. 201806340092 and W.S.H acknowledges support from the European Research Council through ERC CoG 817720 (IMAGINE).

**References**

(1) Rickhaus, P.; Liu, M.-H.; Makk, P.; Maurand, R.; Hess, S.; Zihlmann, S.; Weiss, M.; Richter, K.; Schönenberger, C. Guiding of Electrons in a Few-Mode Ballistic Graphene Channel. *Nano Lett.* **2015**, *15*, 5819–5825.
(2) Williams, J. R.; Low, T.; Lundstrom, M. S.; Marcus, C. M. Gate-Controlled Guiding of Electrons in Graphene. *Nature Nanotech* **2011**, *6*, 222–225.

(3) Sorbello, R. S.; Chu, C. S. Residual Resistivity Dipoles, Electromigration, and Electronic Conduction in Metallic Microstructures. *IBM Research and Development* **1988**, *32*, 58–62.

(4) Zwerger, W.; Bonig, L.; Schonhammer, K. Exact Scattering Theory for the Landauer Residual-Resistivity Dipole. *Phys Rev B Condens Matter* **1991**, *43*, 6434–6439.

(5) Somanchi, S.; Terrés, B.; Peiro, J.; Staggenborg, M.; Watanabe, K.; Taniguchi, T.; Beschoten, B.; Stampfer, C. From Diffusive to Ballistic Transport in Etched Graphene Constrictions and Nanoribbons. *ANNALEN DER PHYSIK* **2017**, *529*, 1700082.

(6) Di Ventra, M. *Electrical Transport in Nanoscale Systems*; Cambridge University Press: Cambridge, 2008.

(7) Datta, S. *Lessons from Nanoelectronics: A New Perspective on Transport — Part A: Basic Concepts*, 2nd ed.; World Scientific, 2017; Vol. 05.

(8) de Jong, M. J. M. Transition from Sharvin to Drude Resistance in High-Mobility Wires. *Phys. Rev. B* **1994**, *49*, 7778–7781.

(9) Lipperheide, R.; Weis, T.; Wille, U. Generalized Drude Model: Unification of Ballistic and Diffusive Electron Transport. *J. Phys.: Condens. Matter* **2001**, *13*, 3347.

(10) Csontos, D.; Ulloa, S. E. Crossover from Diffusive to Quasi-Ballistic Transport. *Journal of Applied Physics* **2007**, *101*, 033711.

(11) Geng, H.; Deng, W.-Y.; Ren, Y.-J.; Sheng, L.; Xing, D.-Y. Unified Semiclassical Approach to Electronic Transport from Diffusive to Ballistic Regimes*. *Chinese Phys. B* **2016**, *25*, 097201.

(12) Fenton, E. W. Electric-Field Conditions for Landauer and Boltzmann-Drude Conductance Equations. *Phys. Rev. B* **1992**, *46*, 3754–3770.

(13) Pastawski, H. M. Classical and Quantum Transport from Generalized Landauer-B\"uttiker Equations. *Phys. Rev. B* **1991**, *44*, 6329–6339.

(14) Borunda, M. F.; Hennig, H.; Heller, E. J. Ballistic versus Diffusive Transport in Graphene. *Phys. Rev. B* **2013**, *88*, 125415.

(15) Handschin, C.; Makk, P.; Rickhaus, P.; Liu, M.-H.; Watanabe, K.; Taniguchi, T.; Richter, K.; Schönenberger, C. Fabry-Pérot Resonances in a Graphene/hBN Moiré Superlattice. *Nano Lett.* **2017**, *17*, 328–333.

(16) Baringhaus, J.; Ruan, M.; Edler, F.; Tejeda, A.; Sicot, M.; Taleb-Ibrahimi, A.; Li, A.-P.; Jiang, Z.; Conrad, E. H.; Berger, C.; Tegenkamp, C.; de Heer, W. A. Exceptional Ballistic Transport in Epitaxial Graphene Nanoribbons. *Nature* **2014**, *506*, 349–354.

(17) White, K. L.; Umantsev, M. A.; Low, J. D.; Custer, J. P. Jr.; Cahoon, J. F. Influence of Geometry on Quasi-Ballistic Behavior in Silicon Nanowire Geometric Diodes. *ACS Appl. Nano Mater.* **2023**, *6*, 5117–5126.

(18) Chang, K.; Eichler, A.; Rhensius, J.; Lorenzelli, L.; Degen, C. L. Nanoscale Imaging of Current Density with a Single-Spin Magnetometer. *Nano Lett* **2017**, *17*, 2367–2373.

(19) Sulpizio, J. A.; Ella, L.; Rozen, A.; Birkbeck, J.; Perello, D. J.; Dutta, D.; Ben-Shalom, M.; Taniguchi, T.; Watanabe, K.; Holder, T.; Queiroz, R.; Principi, A.; Stern, A.; Scaffidi, T.; Geim, A. K.; Ilani, S. Visualizing Poiseuille Flow of Hydrodynamic Electrons. *Nature* **2019**, *576*, 75–79.

(20) Palm, M. L.; Huxter, W. S.; Welter, P.; Ernst, S.; Scheidegger, P. J.; Diesch, S.; Chang, K.; Rickhaus, P.; Taniguchi, T.; Watanabe, K.; Ensslin, K.; Degen, C. L. Imaging of Submicroampere Currents in Bilayer Graphene Using a Scanning Diamond Magnetometer. *Phys. Rev. Applied* **2022**, *17*, 054008.

(21) Finkler, A.; Vasyukov, D.; Segev, Y.; Ne'eman, L.; Lachman, E. O.; Rappaport, M. L.; Myasoedov, Y.; Zeldov, E.; Huber, M. E. Scanning Superconducting Quantum Interference Device on a Tip for Magnetic Imaging of Nanoscale Phenomena. *Review of Scientific Instruments* **2012**, *83*, 073702.

(22) Halbertal, D.; Ben Shalom, M.; Uri, A.; Bagani, K.; Meltzer, A. Y.; Marcus, I.; Myasoedov, Y.; Birkbeck, J.; Levitov, L. S.; Geim, A. K.; Zeldov, E. Imaging Resonant Dissipation from Individual Atomic Defects in Graphene. *Science* **2017**, *358*, 1303–1306.

(23) Ella, L.; Rozen, A.; Birkbeck, J.; Ben-Shalom, M.; Perello, D.; Zultak, J.; Taniguchi, T.; Watanabe, K.; Geim, A. K.; Ilani, S.; Sulpizio, J. A. Simultaneous Voltage and Current Density Imaging of Flowing Electrons in Two Dimensions. *Nat Nanotechnol* **2019**, *14*, 480–487.

(24) Druga, T.; Wenderoth, M.; Homoth, J.; Schneider, M. A.; Ulbrich, R. G. A Versatile High Resolution Scanning Tunneling Potentiometry Implementation. *The Review of scientific instruments* **2010**, *81*, 083704.

(25) Marković, T.; Huang, W.; Gambardella, P.; Stepanow, S. Performance Analysis and Implementation of a Scanning Tunneling Potentiometry Setup: Toward Low-Noise and High-Sensitivity Measurements of the Electrochemical Potential. *Review of Scientific Instruments* **2021**, *92*, 103707.

(26) Lüpke, F.; Korte, S.; Cherepanov, V.; Voigtländer, B. Scanning Tunneling Potentiometry Implemented into a Multi-Tip Setup by Software. *Review of Scientific Instruments* **2015**, *86*, 123701.

(27) Ji, S. H.; Hannon, J. B.; Tromp, R. M.; Perebeinos, V.; Tersoff, J.; Ross, F. M. Atomic-Scale Transport in Epitaxial Graphene. *Nature materials* **2012**, *11*, 114–119.

(28) Willke, P.; Kotzott, T.; Pruschke, T.; Wenderoth, M. Magnetotransport on the Nano Scale. *Nat Commun* **2017**, *8*, 15283.

(29) Sinterhauf, A.; Traeger, G. A.; Momeni Pakdehi, D.; Schadlich, P.; Willke, P.; Speck, F.; Seyller, T.; Tegenkamp, C.; Pierz, K.; Schumacher, H. W.; Wenderoth, M. Substrate Induced Nanoscale Resistance Variation in Epitaxial Graphene. *Nat Commun* **2020**, *11*, 555.

(30) Willke, P.; Druga, T.; Ulbrich, R. G.; Schneider, M. A.; Wenderoth, M. Spatial Extent of a Landauer Residual-Resistivity Dipole in Graphene Quantified by Scanning Tunnelling Potentiometry. *Nature communications* **2015**, *6*, 6399.

(31) Clark, K. W.; Zhang, X.-G.; Vlassiouk, I. V.; He, G.; Feenstra, R. M.; Li, A.-P. Spatially Resolved Mapping of Electrical Conductivity across Individual Domain (Grain) Boundaries in Graphene. *ACS Nano* **2013**, *7*, 7956–7966.

(32) Clark, K. W.; Zhang, X. G.; Gu, G.; Park, J.; He, G.; Feenstra, R. M.; Li, A.-P. Energy Gap Induced by Friedel Oscillations Manifested as Transport Asymmetry at Monolayer-Bilayer Graphene Boundaries. *Physical Review X* **2014**, *4*.

(33) Sinterhauf, A.; Traeger, G. A.; Momeni, D.; Pierz, K.; Schumacher, H. W.; Wenderoth, M. Unraveling the Origin of Local Variations in the Step Resistance of Epitaxial Graphene on SiC: A Quantitative Scanning Tunneling Potentiometry Study. *Carbon* **2021**, *184*, 463–469.

(34) Wang, W.; Munakata, K.; Rozler, M.; Beasley, M. R. Local Transport Measurements at Mesoscopic Length Scales Using Scanning Tunneling Potentiometry. *Physical Review Letters* **2013**, *110*.

(35) Zhou, X.; Ji, S.-H.; Chockalingam, S. P.; Hannon, J. B.; Tromp, R. M.; Heinz, T. F.; Pasupathy, A. N.; Ross, F. M. Electrical Transport across Grain Boundaries in Graphene Monolayers on SiC(0 0 0 $\bar{1}$). *2D Mater.* **2018**, *5*, 031004.

(36) De Cecco, A.; Prudkovskiy, V. S.; Wander, D.; Ganguly, R.; Berger, C.; de Heer, W. A.; Courtois, H.; Winkelmann, C. B. Non-Invasive Nanoscale Potentiometry and Ballistic Transport in Epigraphene Nanoribbons. *Nano Lett* **2020**, *20*, 3786–3790.

(37) Krebs, Z. J.; Behn, W. A.; Li, S.; Smith, K. J.; Watanabe, K.; Taniguchi, T.; Levchenko, A.; Brar, V. W. Imaging the Breaking of Electrostatic Dams in Graphene for Ballistic and Viscous Fluids. *Science* **2023**, *379*, 671–676.

(38) Landauer, R. Residual Resistivity Dipoles. *Z Physik B* **1975**, *21*, 247–254.

(39) Landauer, R. Spatial Variation of Currents and Fields Due to Localized Scatterers in Metallic Conduction. *IBM Journal of Research and Development* **1957**, *1*, 223–231.

(40) Landauer, R. Spatial Variation of Currents and Fields Due to Localized Scatterers in Metallic Conduction. *IBM Research and Development* **1988**, *32*, 306–336.

(41) Wang, W.; Beasley, M. R. Measurement of Specific Contact Resistivity Using Scanning Voltage Probes. *Applied Physics Letters* **2013**, *102*.

(42) Lucas, A. Stokes Paradox in Electronic Fermi Liquids. *Phys. Rev. B* **2017**, *95*, 115425.

(43) Weingart, S.; Bock, C.; Kunze, U.; Speck, F.; Seyller, Th.; Ley, L. Low-Temperature Ballistic Transport in Nanoscale Epitaxial Graphene Cross Junctions. *Appl. Phys. Lett.* **2009**, *95*, 262101.

(44) Wang, Q.; Zhang, W.; Wang, L.; He, K.; Ma, X.; Xue, Q. Large-Scale Uniform Bilayer Graphene Prepared by Vacuum Graphitization of 6H-SiC(0001) Substrates. *Journal of Physics: Condensed Matter* **2013**, *25*.

(45) Hannon, J. B.; Tromp, R. M. Pit Formation during Graphene Synthesis on SiC(0001):In Situelectron Microscopy. *Physical Review B* **2008**, *77*.

(46) Sandin, A.; Rowe, J. E.; Dougherty, D. B. Improved Graphene Growth in UHV: Pit-Free Surfaces by Selective Si Etching of SiC(0001)–Si with Atomic Hydrogen. *Surface Science* **2013**, *611*, 25–31.

(47) Rutter, G. M.; Guisinger, N. P.; Crain, J. N.; Jarvis, E. A. A.; Stiles, M. D.; Li, T.; First, P. N.; Stroscio, J. A. Imaging the Interface of Epitaxial Graphene with Silicon Carbide via Scanning Tunneling Microscopy. *Physical Review B* **2007**, *76*, 235416.

(48) Lauffer, P.; Emtsev, K. V.; Graupner, R.; Seyller, T.; Ley, L.; Reshanov, S. A.; Weber, H. B. Atomic and Electronic Structure of Few-Layer Graphene on SiC(0001) Studied with Scanning Tunneling Microscopy and Spectroscopy. *Physical Review B* **2008**, *77*, 155426.

(49) Riedl, C.; Coletti, C.; Starke, U. Structural and Electronic Properties of Epitaxial Graphene on SiC(0 0 0 1): A Review of Growth, Characterization, Transfer Doping and Hydrogen Intercalation. *Journal of Physics D: Applied Physics* **2010**, *43*.

(50) Bolen, M. L.; Harrison, S. E.; Biedermann, L. B.; Capano, M. A. Graphene Formation Mechanisms on $4H\text{-SiC}(0001)$. *Phys. Rev. B* **2009**, *80*, 115433.

(51) Sun, G. F.; Liu, Y.; Rhim, S. H.; Jia, J. F.; Xue, Q. K.; Weinert, M.; Li, L. Si Diffusion Path for Pit-Free Graphene Growth on SiC(0001). *Phys. Rev. B* **2011**, *84*, 195455.

(52) Nie, S.; Feenstra, R. M. Tunneling Spectroscopy of Graphene and Related Reconstructions on SiC(0001). *Journal of Vacuum Science & Technology A* **2009**, *27*, 1052–1057.

(53) Jackson, J. D. *Classical Electrodynamics, 3rd Edition*; Wiley: New York, 1998.

(54) Dutreix, C.; Gonzalez-Herrero, H.; Brihuega, I.; Katsnelson, M. I.; Chapelier, C.; Renard, V. T. Measuring the Berry Phase of Graphene from Wavefront Dislocations in Friedel Oscillations. *Nature* **2019**, *574*, 219–222.

(55) Morozov, S. V.; Novoselov, K. S.; Katsnelson, M. I.; Schedin, F.; Elias, D. C.; Jaszczak, J. A.; Geim, A. K. Giant Intrinsic Carrier Mobilities in Graphene and Its Bilayer. *Phys. Rev. Lett.* **2008**, *100*, 016602.
(56) Tanabe, S.; Sekine, Y.; Kageshima, H.; Nagase, M.; Hibino, H. Observation of Band Gap in Epitaxial Bilayer Graphene Field Effect Transistors. *Jpn. J. Appl. Phys.* **2011**, *50*, 04DN04.
(57) Coelho, R. C. V.; Doria, M. M. Lattice Boltzmann Method for Semiclassical Fluids. *Computers & Fluids* **2018**, *165*, 144–159.
(58) Bhatnagar, P. L.; Gross, E. P.; Krook, M. A Model for Collision Processes in Gases. I. Small Amplitude Processes in Charged and Neutral One-Component Systems. *Phys. Rev.* **1954**, *94*, 511–525.
(59) Hammer, R.; Fritz, V.; Bedoya-Martínez, N. The Worm-LBM, an Algorithm for a High Number of Propagation Directions on a Lattice Boltzmann Grid: The Case of Phonon Transport. *International Journal of Heat and Mass Transfer* **2021**, *170*, 121030.
(60) Feenstra, R. M.; Briner, B. G. The Search for Residual Resistivity Dipoles by Scanning Tunneling Potentiometry. *Superlattices and Microstructures* **1998**, *23*, 699–709.
(61) Briner, B. G.; Feenstra, R. M.; Chin, T. P.; Woodall, J. M. Local Transport Properties of Thin Bismuth Films Studied by Scanning Tunneling Potentiometry. *Physical Review B* **1996**, *54*, R5283–R5286.
(62) Lüpke, F.; Eschbach, M.; Heider, T.; Lanius, M.; Schüffelgen, P.; Rosenbach, D.; von den Driesch, N.; Cherepanov, V.; Mussler, G.; Plucinski, L.; Grützmacher, D.; Schneider, C. M.; Voigtländer, B. Electrical Resistance of Individual Defects at a Topological Insulator Surface. *Nat Commun* **2017**, *8*, 15704.
(63) Huang, W.; Seo, J. A.; Canavan, M. P.; Gambardella, P.; Stepanow, S. Observation of Different Li Intercalation States and Local Doping in Epitaxial Mono- and Bilayer Graphene on SiC(0001). *Nanoscale* **2024**, *16*, 3160–3165.
(64) Palm, M. L.; Ding, C.; Huxter, W. S.; Taniguchi, T.; Watanabe, K.; Degen, C. L. Observation of Current Whirlpools in Graphene at Room Temperature. *Science* **2024**, *384*, 465–469.
(65) Kotov, V. N.; Uchoa, B.; Pereira, V. M.; Guinea, F.; Castro Neto, A. H. Electron-Electron Interactions in Graphene: Current Status and Perspectives. *Reviews of Modern Physics* **2012**, *84*, 1067–1125.
(66) Polini, M.; Geim, A. K. Viscous Electron Fluids. *Physics Today* **2020**, *73*, 28–34.